\documentclass[twocolumn,superscriptaddress]{revtex4}
\usepackage{graphicx,amsmath,amssymb,amsthm}

\def\textbf#1{{\bf #1}}
\def\be{\begin{equation}}
\def\ee{\end{equation}}
\def\ben{\begin{eqnarray}}
\def\een{\end{eqnarray}}
\def\eea{\end{array}}
\def\bea{\begin{array}}

\newcommand{\tr}[1]{\mathrm{Tr}#1}
\newcommand{\bei}{\begin{itemize}}
\newcommand{\eei}{\end{itemize}}
\newcommand{\ket}[1]{|#1\rangle}
\newcommand{\bra}[1]{\langle#1|}

\newcommand{\proj}[1]{\ket{#1}\!\bra{#1}}

\newcommand{\eg}{{\it{e.g.~}}}
\newcommand{\ie}{{\it{i.e.~}}}
\newcommand{\etal}{{\it{et al.}}}

\begin{document}

\title{Quantum networks reveal quantum nonlocality}

\author{D. Cavalcanti}
\affiliation{Centre for Quantum Technologies, National University
of Singapore, 117542 Singapore }

\author{M. L. Almeida}
\affiliation{ICFO-Institut de Ciencies Fotoniques, 08860
Castelldefels (Barcelona), Spain}

\author{V. Scarani}
\affiliation{Centre for Quantum Technologies, National University
of Singapore, 117542 Singapore } \affiliation{Department of
Physics, National University of Singapore, 117542 Singapore}

\author{A. Ac\'\i n}
 \affiliation{ICFO-Institut de Ciencies Fotoniques, 08860 Castelldefels (Barcelona), Spain}
 \affiliation{ICREA-Institucio Catalana de Recerca i Estudis Avan\c cats, 08010 Barcelona, Spain}

\begin{abstract}
The results of local measurements on some composite quantum
systems cannot be reproduced classically. This
impossibility, known as quantum nonlocality, represents a
milestone in the foundations of quantum theory. Quantum
nonlocality is also a valuable resource for information processing
tasks, \eg quantum communication, quantum key
distribution, quantum state
estimation, or randomness extraction. Still, deciding if a quantum state is
nonlocal remains a challenging problem. Here we introduce a novel
approach to this question: we study the nonlocal properties of
quantum states when distributed and measured in networks. Using
our framework, we show how any one-way entanglement distillable
state leads to nonlocal correlations. Then, we prove that
nonlocality is a non-additive resource, which can be activated.
There exist states, local at the single-copy level, that become
nonlocal when taking several copies of it. Our results imply that
the nonlocality of quantum states strongly depends on the
measurement context.
\end{abstract}

\maketitle

Nonlocality is a property of the outcome distributions resulting
from local measurements on composite physical systems. Consider a
system shared by $N$ parties, which perform $m$ spacelike
separated measurements with $r$ possible results, on their
respective subsystems. Denote by $x_i=1,\ldots,m$ the measurement
chosen by party $i$ and by $a_i=1,\ldots,r$ the corresponding
outcome. The obtained joint probability distribution
$P(a_1,\ldots,a_N|x_1,\ldots,x_N)$ is local whenever it can be
explained as the result of classically correlated data,
represented by $\lambda$, following a distribution $p(\lambda)$,
\ie
\begin{multline}\label{lcorr}
    P_\mathrm{local}(a_1,\ldots,a_N|x_1,\ldots,x_N)=\\
    =\sum_\lambda p(\lambda) P(a_1|x_1,\lambda)\ldots
    P(a_N|x_N,\lambda)\,.
\end{multline}
Every local distribution~\eqref{lcorr}
satisfies linear constraints known as Bell inequalities. The
violation of a Bell inequality is then a signature of
nonlocality: distributions not fitting the description~\eqref{lcorr} are called nonlocal.

Remarkably, local measurements
on some quantum states lead to nonlocal correlations~\cite{bell}.
This can be observed in the original Bell test scenario~\cite{bell}, where the
measurements are performed on a single copy of some quantum state
$\rho$. In other words, there exist quantum
distributions of outcomes
\begin{equation}\label{qcorr}
    P_\mathrm{Q}(a_1,\ldots,a_N|x_1,\ldots,x_N)=\tr(\rho M_{a_1}^{x_1}\otimes\cdots\otimes M_{a_N}^{x_N}) ,
\end{equation}
where $M_{a_i}^{x_i}$ are the positive operators defining the
local measurements by party $i$,
$\sum_{a_i}M_{a_i}^{x_i}=1,\,\forall x_i$, which cannot be
represented by any local model~\eqref{lcorr}. The quantum state
$\rho$ is then called nonlocal. Contrary, if the results of all
local measurements on $\rho$ can be written as in \eqref{lcorr},
it is local at the single-copy level.

A stronger version of nonlocality is possible in a multipartite
scenario. Consider the case in which only a subset of the $N$
parties share nonlocal correlations. Although these correlations
are nonlocal, they are not genuine $N$-partite nonlocal. A quantum
state is said to be genuine $N$-partite nonlocal only when there
exist measurements on it establishing nonlocal correlations among
all the $N$ parties (see also Appendix A).

Deciding whether a quantum state is nonlocal represents not only a
fundamental question; it also has important practical
applications. The success of information-processing applications
as quantum communication complexity~\cite{comm},
no-signalling~\cite{bhk} and device-independent~\cite{deviceind}
quantum key distribution, device-independent quantum state
estimation~\cite{devindstate,McKague}, or randomness extraction
\cite{colbeck,pironio}, crucially relies on the existence of
nonlocal correlations. Unfortunately, identifying the nonlocal
properties of even the simplest families of entangled quantum
state remains an extremely difficult problem
\cite{werner,acin-gisin-toner,barrett, mafs}.

Here we introduce a new framework to the study of quantum
nonlocality. Given an $N$-partite quantum state $\rho$,
the main idea is to create a network of $n\geq N$ parties and to
distribute among them $L$ copies of $\rho$, according to a given
spatial configuration (see Figure 1). If nonlocal correlations
are observed in the network, they can only come from the quantum
state $\rho$ and, therefore, it is called a nonlocal resource.
Within this new scenario, we prove that all one-way entanglement
distillable states are nonlocal resources. Then, we provide
examples of activation of quantum nonlocality, namely quantum
states that are local at the single-copy level but which become
nonlocal in a network scenario or simply by taking several copies
of it. We show that similar activation phenomena occur for genuine
$n$-partite nonlocality.


\begin{figure*}
\includegraphics[width=1.3\columnwidth,keepaspectratio]{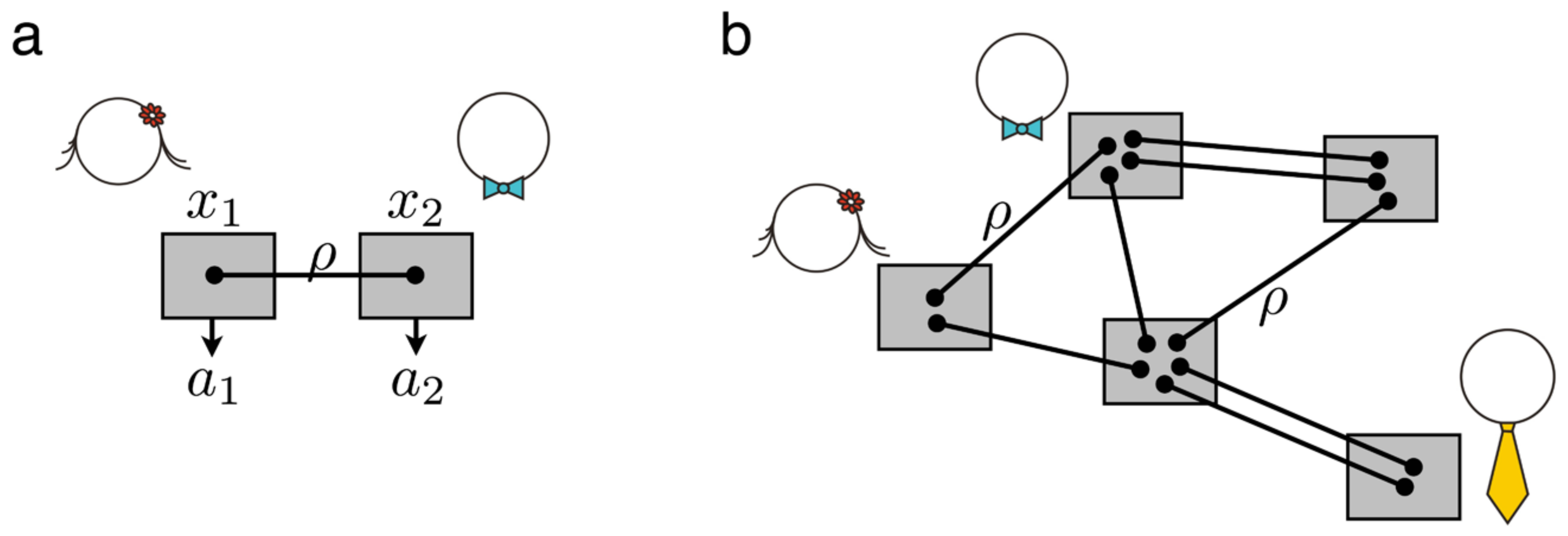}
\caption{\textbf{Nonlocality in the network scenario.}
\textbf{a.} The standard scenario for the study of quantum
nonlocality consists of $N$ parties sharing one copy of an $N$-party quantum state $\rho$, to which they apply $m$ different measurements of $r$ possible results. The measurement choices and corresponding results by the two parties are labeled by $x_1,\ldots,x_N=1,\ldots,m$ and $a_1,\ldots,a_N=1,\ldots,r$. By repeating this process, the parties can estimate the joint probability distribution $P(a_1, \ldots, a_N|x_1,\ldots, x_N)$ describing the measurement statistics and test its nonlocality.
\textbf{b.} We introduce a new scenario, where many copies of the quantum state
$\rho$ are distributed among $n$ parties according to a network configuration.
The parties measure their respective subsystems and check if the
obtained multipartite probability distribution is local, \ie if it can be
written as in \eqref{lcorr}. As shown in the main text, there are
quantum states that are local according to scenario
\textbf{a}, but provide nonlocal correlations in
scenario \textbf{b}. } \label{scenario}
\end{figure*}

There exist two main reasons for our network scenario to offer new
possibilities in the study of quantum nonlocality. First, it enlarges the standard scenario in which a single-copy of an $N$-partite state is given to $N$ observers.
Second, the network scenario allows considering
protocols in which a subset of the parties projects the remaining
ones into a nonlocal quantum state. Indeed, post-selection is not
a valid operation in standard bipartite Bell tests, as it is
associated with sending information between the parties. Nevertheless,
it is now allowed because it is performed on the results of
measurements that are spacelike separated from the measurements
actually used in the Bell test. (Spacelike
separation here plays the role of time-ordering in the
\emph{hidden nonlocality} framework~\cite{hiddenNL,Masanes}).
These ideas are behind the two main technical observations used
next. Consider an $N$-partite quantum state
$\rho$.

\emph{Observation 1.} If there exist local measurements by $k$ parties such
that, for one of the measurement outcomes, the resulting state
among the remaining $N-k$ parties is nonlocal, then the initial
state $\rho$ is necessarily nonlocal.

This fact was first used in Ref.~\cite{pr} to prove that all
multipartite entangled pure states are nonlocal. In Appendix $B$
we provide a
simple proof of Observation 1. 

\emph{Observation 2.} If, for every bipartition of the parties, there exist local
measurements on $N-2$ parties that, for every outcome, create a
maximally-entangled state between parties belonging to the
different partitions, then the initial state $\rho$ is genuine
$N$-partite nonlocal.

This was proven in Ref.~\cite{mafs2} and is actually stronger as
it also implies that the state $\rho$ contains only genuine
$N$-partite nonlocal correlations.

\begin{figure}
\includegraphics[width=0.7\columnwidth,keepaspectratio]{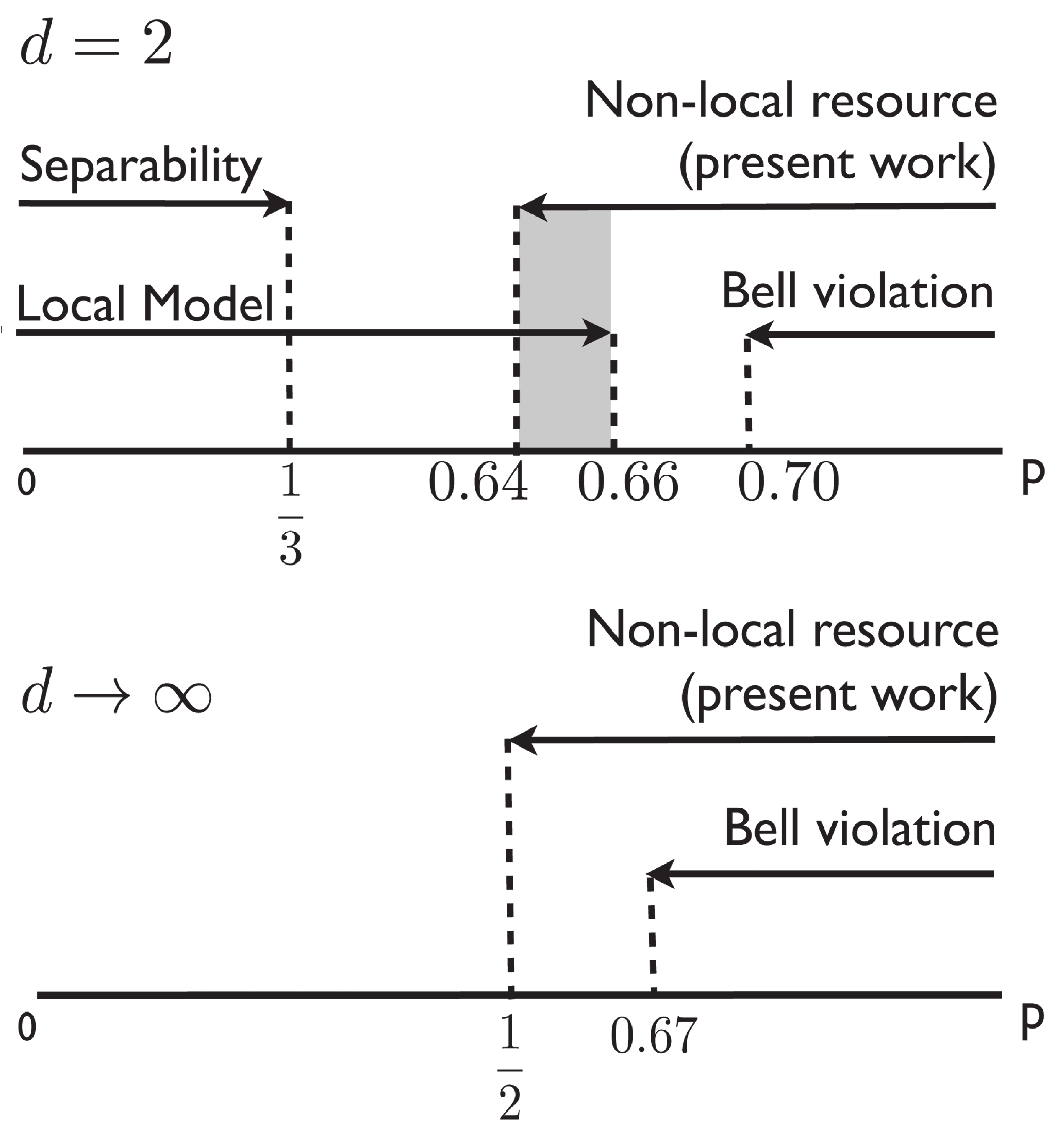}
\caption{\textbf{Nonlocality of isotropic states.} Isotropic states are mixtures of a maximally-entangled state and white noise, see  \eqref{iso}. In the qubit case, $d=2$, these
states violate a Bell inequality for
$p\gtrsim 0.705$~\cite{vertesi}. A local model for any experiment
involving von Neumann measurements on a single copy of these states is
possible for $p\lesssim0.66$~\cite{acin-gisin-toner}. This limit
also holds if general two-outcome measurements are considered (see
Appendix D). In the case of general measurements, the existence of a local model
has been proven for $p\leq5/12$~\cite{barrett}. Here we
show, using two-outcome measurements, that two-qubit isotropic states are
nonlocal in a network scenario for $p\gtrsim0.64$. In the limit of very large dimension,
$d\rightarrow \infty$, isotropic states violate the
Collins-Gisin-Linden-Masar-Popescu (CGLMP) Bell
inequality~\cite{CGLMP} for $d\gtrsim0.67$. We show that these states are nonlocal resources
in our network scenario for $p>1/2$.} \label{bounds}
\end{figure}

Our techniques are fully general and apply to any state. However,
in what follows, we frequently illustrate their usefulness in networks composed of isotropic states in $d\times d$ systems; namely mixtures of maximally-entangled states, $\ket{\Phi}$, and white noise, weighted by the noise parameter $p$,
\begin{equation}\label{iso}
    \rho_{\mathrm I}(p)=p\,\proj{\Phi}+(1-p)\frac{1}{d^2} .
\end{equation}
The known nonlocal properties of these states are summarized in Figure \ref{bounds}.


\begin{figure}\centering
\includegraphics[width=0.6\columnwidth,keepaspectratio]{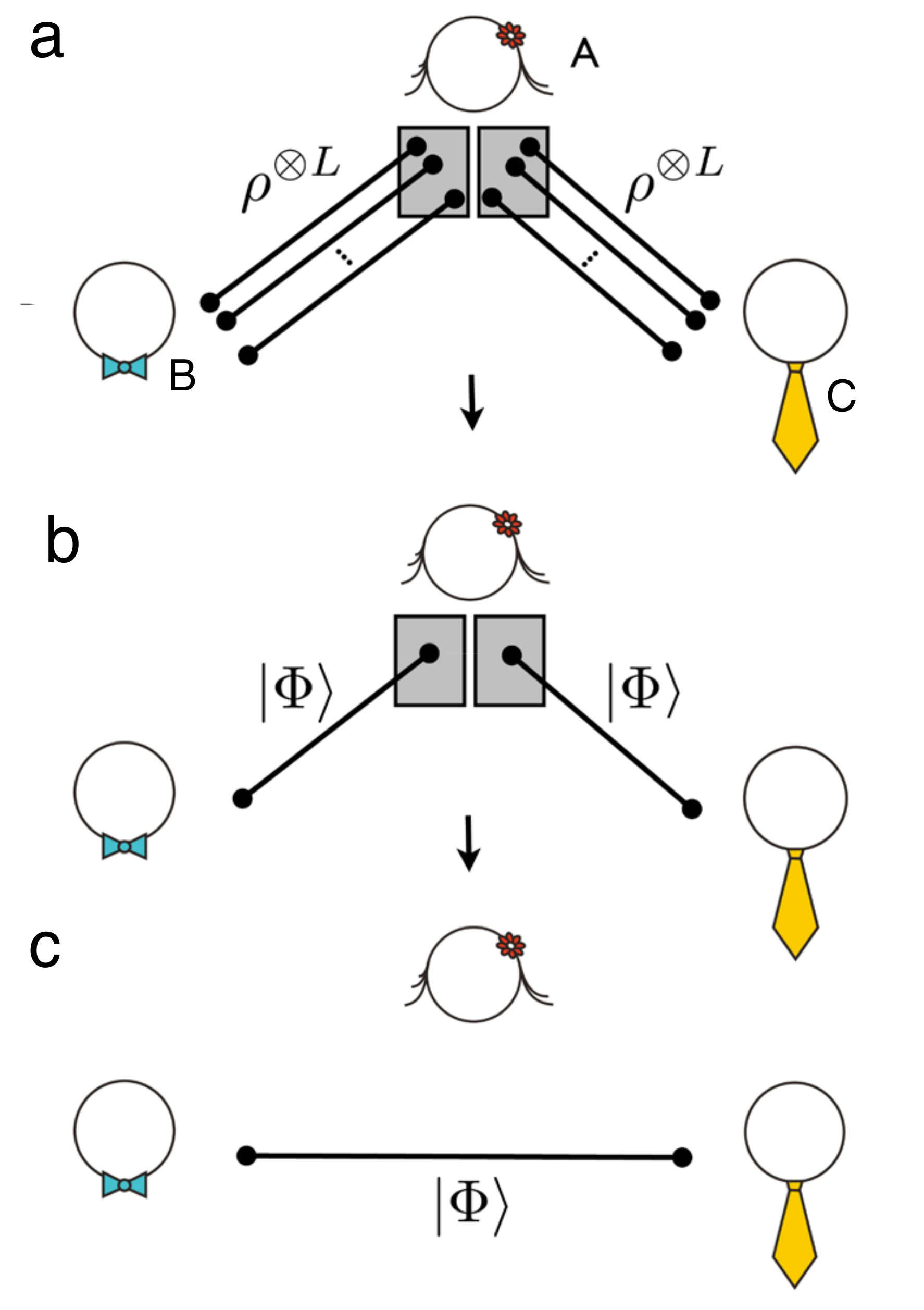}
\caption{\textbf{Any one-way entanglement distillable state is a
nonlocal resource.} \textbf{a.} Consider a network consisting of
three distant parties Alice, Bob, and Charlie. Alice shares $L$
copies of an entangled state with Bob, $\rho_{AB}$, and $L$ copies
of the same state with Charlie, ${\rho}_{AC}$ . \textbf{b.} Assume
this state has non-zero one-way distillable entanglement. Then,
there exists a projection on Alice's part of $\rho_{AB}^{\otimes
L}$ such that the resulting state between Alice and Bob,
$\sigma_{AB}$ , tends to a maximally-entangled state, in the limit
of an infinite number of copies, \ie
$\bra{\Phi}\sigma_{AB}\ket{\Phi}\rightarrow 1$ when
$L\rightarrow\infty$. The same projection is applied by Alice to
her part of $\rho_{AC}^{\otimes L}$. \textbf{c.} Alice projects
now her subsystems into a maximally-entangled state. This, in
turn, projects Bob and Charlie into a state which is arbitrarily
close to a maximally-entangled state. Clearly, all the previous
projections by Alice can be seen as a one-shot measurement that,
with non-zero probability, leaves Bob and Charlie in a state
arbitrarily close to a maximally-entangled state. Since this state
is nonlocal, we conclude that the network state is also nonlocal
(Observation 1). In the case of isotropic states \eqref{iso}, this
procedure allows the detection of nonlocality for $p>1/2$ when
$d\rightarrow\infty$.
This value is obtained by computing the hashing bound, which is a lower bound on the one-way distillable entanglement~\cite{devetak-winter}, of these states.} 
\label{one-way}
\end{figure}

We start by showing that any one-way entanglement distillable states form a quantum network that
displays nonlocal correlations. The required network consists of
three parties in a $\Lambda$ configuration (see Figure
\ref{one-way}): one of the parties, say Alice, shares $L$ copies
of $\rho$ with Bob, and $L$ different copies with Charlie. If
$\rho$ is one-way entanglement distillable, there exists a
measurement result by Alice which projects Bob and Charlie into a
state that can be made, by increasing $L$, arbitrarily close to a
maximally-entangled state. As this state is nonlocal, the network
state must be nonlocal by virtue of Observation 1. As an illustration,
consider a network made of isotropic states \eqref{iso}. In the
limit of large dimension, nonlocal correlations are observed
whenever $p>1/2$, which is far beyond the known region of Bell
violation for the single-copy $\rho_{\mathrm I}$ (see Figure~\ref{bounds}).


We now move to the announced examples of activation of quantum
nonlocality. The first example is concerned with genuine
multipartite nonlocality and is remarkably simple. Consider a
bipartite maximally-entangled state $\ket{\Phi}$, which clearly is not genuine tripartite nonlocal. However, two copies of $\ket{\Phi}$ disposed in a $\Lambda$
configuration, as above, compose a network with genuine tripartite
nonlocality. The network state reads $\ket{\Phi}_{AB}\otimes\ket{\Phi}_{A'C}$, which is
tripartite nonlocal according to Observation 2 (actually, it
is fully tripartite nonlocal \cite{mafs2}). This example can
be used to construct a more standard example of activation, where the tripartite quantum nonlocality of a state is activated
by the state itself. The state reads
\begin{multline}\label{act3state}
    \sigma=\frac{1}{2}(\proj{\psi_1}_{ABC}\otimes\proj{000}_{A_fB_fC_f}+\\
    +\proj{\psi_2}_{ABC}\otimes\proj{111}_{A_fB_fC_f}) ,
\end{multline}
where $\ket{\psi_1}_{ABC}=\ket{\Phi}_{AB}\otimes\ket{\phi}_C$ and
$\ket{\psi_2}_{ABC}=\ket{\Phi}_{AC}\otimes\ket{\phi}_B$, being
$\ket{\phi}$ some arbitrary state. Now, although $\sigma$ does not
contain any genuine tripartite nonlocality, $\sigma^{\otimes L}$
becomes genuine tripartite nonlocal for large $L$ (see Appendix
C). Such results can be generalized to an arbitrary number of
parties: a star-shape $n$-party network, in which the central node
is connected to the remaining nodes by one copy of a state
$\ket{\Phi}$, is genuine $n$-partite nonlocal.


\begin{figure}\centering
\includegraphics[width=0.7\columnwidth,keepaspectratio]{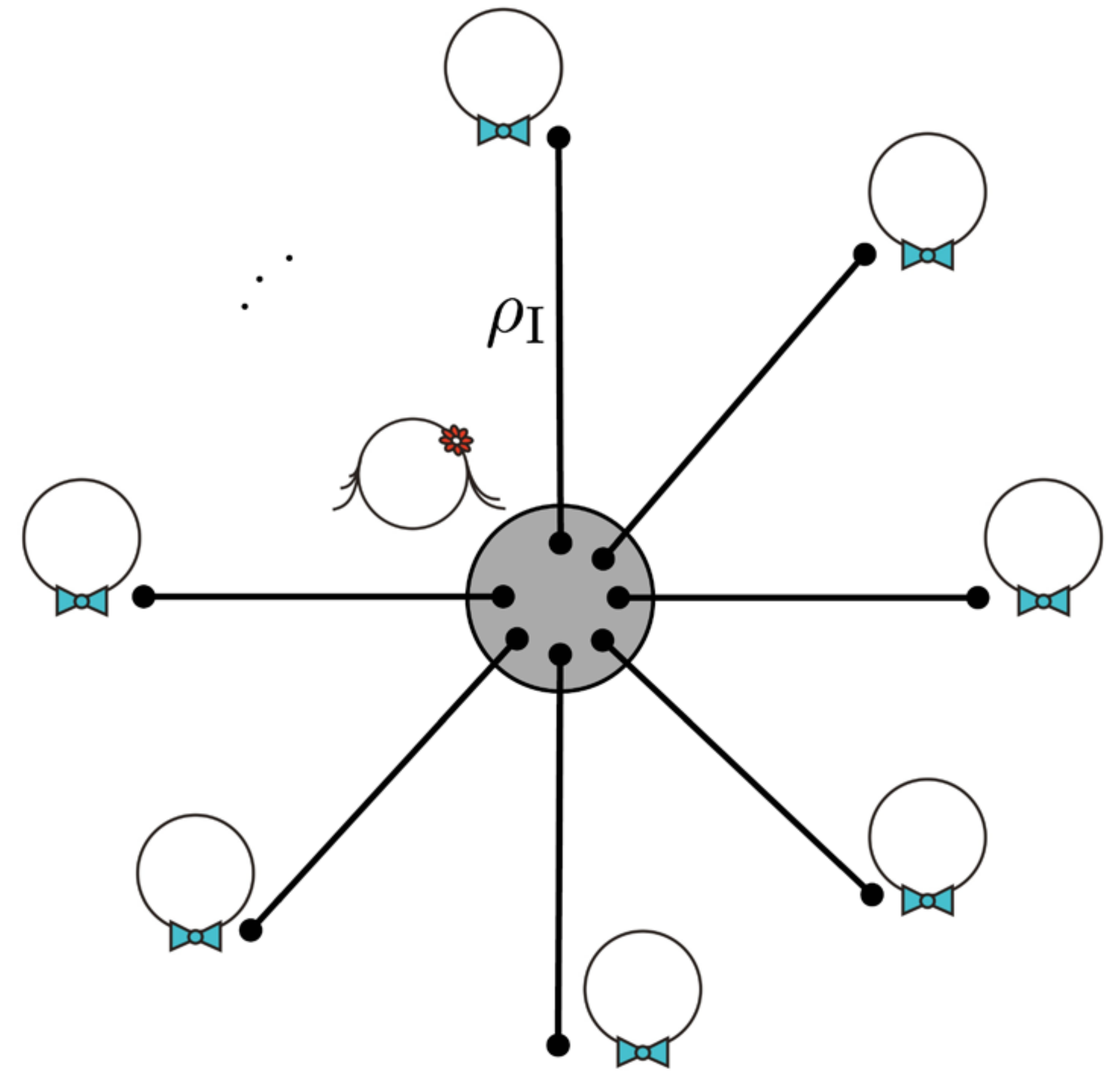}
\caption{\textbf{Activation of nonlocality.} Consider the network
scenario in which $N+1$ parties share two-qubit isotropic states
$\rho_{\mathrm{I}}$ in a star-like configuration. If
the central node, Alice, projects her qubits into the state
$\ket{GHZ}=(\ket{0...0}+\ket{1...1})/\sqrt{2}$, the remaining $N$
parties are left with a state that violates a Bell inequality for
$p>p_N=(2/\pi)2^{1/N}$ \cite{sen}. It then follows from Observation 1
that the network state is nonlocal and $\rho_{\mathrm{I}}$ is a
nonlocal resource in this region. For $N\geq 7$, this bound is
smaller than the noise corresponding to the best known Bell violation~\cite{vertesi}, \ie $p_7<0.705$.
More interestingly, if $N\geq 21$ one has $p_N<0.66$. This is precisely the noise value for which
the existence of a local model for von Neumann measurements on
$\rho_{\mathrm{I}}$ has been proven~\cite{acin-gisin-toner}.
As the local model in Ref.~\cite{acin-gisin-toner} can easily be extended to general two-outcome measurements (see Appendix D) and the previous Bell violation only involved two-outcome measurements, we conclude that the nonlocality of $\rho_{\mathrm{I}}$ is activated by the network configuration. That is, nonlocal correlations are
observed in networks consisting of isotropic states that are local at the single-copy
level.}
\label{star}
\end{figure}

Moving to standard nonlocality, our example of activation uses
isotropic two-qubit states~\eqref{iso} disposed again in a
star-shape network, as depicted in Fig. \ref{star}. In the noise
region $p\gtrsim0.64$, there exists a local measurement result in
the central node that leaves the remaining subsystems in a
nonlocal state~\cite{sen}. Then, Observation 1 guarantees that the
corresponding isotropic states are nonlocal resources. But, when
$p\lesssim 0.66$, these states are known to be local at the
single-copy level for von Neumann measurements
\cite{acin-gisin-toner} (and, consequently, for general dichotomic
measurements, see Appendix D). In the overlap of these regions,
nonlocality is activated: local (single-copy) isotropic states
form a nonlocal quantum network. The minimal size of the nonlocal
network required to enter the single-copy locality region is 21
parties. As in the previous example~\eqref{act3state}, classically
correlated states can be added to construct an example of
activation in which a 21-particle local state $\tau$ becomes
nonlocal by taking a sufficiently large number of copies of it,
$\tau^{\otimes L}$ (see Appendix C).


To conclude, here we show that a much better use of the nonlocal potential of quantum states is achieved simply by distributing them in networks. Our work opens new perspectives in the study of quantum nonlocality. For instance, it would be interesting to analyze if, for any entangled state, there exists a network displaying nonlocal correlations. In fact, despite considerable efforts after Bell's seminal work, we are still far from understanding the exact relation between entanglement and nonlocality. Our results show that this relation is much subtler than initially expected: the nonlocal character of a quantum state strongly depends on its measurement context, namely on the network configuration.

\begin{acknowledgements}
We thank A. Winter and A. Salles for interesting comments. This
work was supported by the European Q-Essence project and ERC
Starting grant PERCENT, the Spanish MEC FIS2007-60182 and
Consolider-Ingenio QOIT projects, Generalitat de Catalunya, Caixa
Manresa, National Research Foundation and the Ministry of
Education of Singapore.
\end{acknowledgements}

%

\bibliographystyle{naturemag}

\clearpage

\section{Appendices}

\subsection{Definition of genuine multipartite nonlocality.}
In this section we provide a formal definition of genuine
multipartite nonlocality, a stronger version of nonlocality that
can be defined in the multipartite scenario \cite{svetlichny,
svetlichnyN,collins,mafs2,stefano}. Consider, for instance, a
situation in which the correlations observed among $N$ parties can
be written as
\begin{multline}\label{locnonlocex}
    P_\mathrm{L:NL}(a_1,\ldots,a_N|x_1,\ldots,x_N)\\=\sum_\lambda p(\lambda) P(a_1,\ldots,a_k|x_1,\ldots,x_k,\lambda)\times\\P(a_{k+1},\ldots,a_N|x_{k+1},\ldots,x_N,\lambda).
\end{multline}
That is, the distribution can be simulated by a hybrid
local/nonlocal model in which nonlocal correlations are given to
the first $k$ parties and to the $N-k$ remaining ones, but
these two groups are correlated only through the classical random variable $\lambda$.
Despite the fact that
hybrid distributions \eqref{locnonlocex} can be nonlocal, in the sense of violating \eqref{lcorr},
they are not genuinely multipartite nonlocal, \ie, the nonlocal
correlations are not shared by all the parties in the system.
This observation naturally leads to the concept of genuine
$N$-partite nonlocality.

A probability distribution is said to be genuine
$N$-partite nonlocal, or contain genuine $N$-partite correlations,
whenever it cannot be reproduced by the
combination of any hybrid models
\eqref{locnonlocex}, that is,
\begin{equation}\label{locnonloc}
    P_\mathrm{MNL}(a_1,\ldots,a_N|x_1,\ldots,x_N)\neq\sum_{\lambda}
    p(\lambda)P^{\lambda}_\mathrm{L:NL}\,.
\end{equation}
Here, the terms $P^\lambda_\mathrm{L:NL}$ are such that there
exists a splitting of the $N$ parties into two groups, which may
depend on $\lambda$, allowing a decomposition like in
\eqref{locnonlocex}. Consequently, an $N$-partite quantum state
is said to be genuine $N$-partite nonlocal whenever there exist local measurement by the parties
leading to genuine $N$-partite nonlocal correlations.

It is important to mention here that, throughout this work, we
always assume the validity of the no-signalling principle. Thus,
all the terms $P^\lambda_\mathrm{L:NL}$ must be compatible with
such principle (see also~\cite{stefano}). This does not coincide
with the original definition of genuine $N$-partite nonlocality
given in Refs.~\cite{svetlichny, svetlichnyN,collins} but has a
natural operational meaning: the parties are assumed not to be
able to transmit information instantaneously even if they have
access to the variable $\lambda$.

Similar to standard nonlocality, genuine multipartite nonlocality
is usually detected by the violation of some linear inequalities,
known as Svetlichny inequalities, that are satisfied by any hybrid
local/nonlocal
distributions~\cite{svetlichny,svetlichnyN,collins}.

\subsection{Constructing Bell inequality for $N$-partite systems}

In this section we prove Observation 1. The goal, then, is to show
that if an $N$-party state $\rho$ is such that
there exist local projections by $k$ parties
mapping the remaining $N-k$ parties into a nonlocal state, $\rho_{N-k}$, then the initial state $\rho$ is also nonlocal. The proof is based on the fact that it is always
possible to construct a Bell inequality violated by the state $\rho$ from
a Bell inequality violated by the post-measurement nonlocal state $\rho_{N-k}$.

For simplicity in the notation, we consider the case in which the
local projections are performed by the first $k$ parties.
Take any Bell inequality violated by the nonlocal state $\rho_{N-k}$,
\begin{equation}\label{Bell ineq1}
\mathcal{I}_{N-k}=\sum_{\mathbf{a},\mathbf{x}}c_{\mathbf{a},\mathbf{x}}P(\mathbf{a}|\mathbf{x})\leq
K\,,
\end{equation}
where $\mathbf{a}=a_{k+1}\ldots a_N$ represent the outcomes of the
measurements $\mathbf{x}=x_{k+1}\ldots x_N$ performed by the
parties $k+1,\ldots,N$, and $c_{\mathbf{a},\mathbf{x}}$ and $K$ are the
weights and local bound defining the inequality. Since the state $\rho_{N-k}$
is obtained for a particular outcome $\mathbf{b'}=b'_{1}\ldots
b'_k$ of local measurements $\mathbf{y'}=y'_{1}\ldots y'_k$ at
parties $1,\ldots,k$ applied to $\rho$, we can
use \eqref{Bell ineq1} to write a generalized Bell
inequality~\cite{pr} violated by $\rho$,
\begin{equation}\label{cond Bell ineq1}
\mathcal{I}_N=\sum_{\mathbf{a},\mathbf{x}}c_{\mathbf{a},\mathbf{x}}P(\mathbf{a}|\mathbf{x},
\mathbf{b'},\mathbf{y'})\leq K\,.
\end{equation}
Generalized Bell inequalities differentiate from standard ones in
the sense that they consider distributions conditioned on
particular measurement outcomes, \ie they assume post-selection of
events.

In general, it is not possible to transform a generalized Bell
inequality into a standard Bell inequality. However, this turns
out to be possible in our case because the local measurements and
results corresponding to the post-selection, $\mathbf{y'}$ and
$\mathbf{b'}$, define events that are spacelike separated from the
measurements appearing in the Bell inequality \eqref{Bell ineq1},
$\mathbf{x}$. The no-signalling condition then guarantees that the
event $(\mathbf{b'},\mathbf{y'})$ can be written independently
from measurements $\mathbf{x}$
\begin{equation}\label{factor prob}
P(\mathbf{a},\mathbf{b'}|\mathbf{x},
\mathbf{y'})=P(\mathbf{b'}|\mathbf{y'})P(\mathbf{a}|\mathbf{x},\mathbf{b'},
\mathbf{y'})\,.
\end{equation}
Using this, it directly follows from inequality~\eqref{cond Bell ineq1} that
the $N$-partite state $\rho$ must violate
\begin{equation}\label{cond Bell ineq2}
\mathcal{I}_N=\sum_{\mathbf{a},\mathbf{x}}c_{\mathbf{a},\mathbf{x}}P(\mathbf{a},\mathbf{b'}|\mathbf{x},\mathbf{y'})\leq
K P(\mathbf{b'}|\mathbf{y'})\,,
\end{equation}
which is a standard Bell inequality satisfied by all local models~\eqref{lcorr}.

\subsection{Activation of nonlocality in the many-copy scenario.}

The purpose of this section is to show how the examples of
activation described in the main text,  which are based on network
configurations, can be mapped into more standard examples of
activation, in which the nonlocal properties of a quantum state
change by taking copies of it. The main idea is to provide all the
parties with classically correlated \emph{flags} which allow them
to reconstruct the activation network, with a probability that can
be made arbitrarily close to one by increasing the number of
copies. We illustrate this procedure in the simplest example of
activation of genuine multipartite nonlocality, but the same
construction can be easily applied to the example of activation of
standard nonlocality.

Consider the tripartite state
\begin{multline}\label{act3state2}
    \sigma=\frac{1}{2}(\proj{\psi_1}_{ABC}\otimes\proj{000}_{A_fB_fC_f}\\+
    \proj{\psi_2}_{ABC}\otimes\proj{111}_{A_fB_fC_f}) ,
\end{multline}
introduced in the main text, where
$\ket{\psi_1}_{ABC}=\ket{\Phi}_{AB}\otimes\ket{\phi}_C$ and
$\ket{\psi_2}_{ABC}=\ket{\Phi}_{AC}\otimes\ket{\phi}_B$, being
$\ket{\phi}$ some arbitrary state. Clearly this state has only
bipartite nonlocal correlations. We show in what follows that
$\sigma^{\otimes L}$, for sufficiently large $L$, is genuine
tripartite nonlocal. As said, in order to do that it is convenient
to interpret the qubits in systems $A_f$, $B_f$ and $C_f$ as
flags. Moreover, it is important to recall that two bipartite
maximally-entangled states in a $\Lambda$ configuration,
$\ket{\Psi}=\ket{\Phi}_{AB}\ket{\Phi}_{A'C}$, define a genuine
tripartite nonlocal state according to Observation 2. This means
that there exist local measurements by the parties, $M_x^{a}$,
$M_y^{b}$ and $M_z^{c}$, such that the corresponding correlations,
described by
\begin{equation}\label{prob_Psi}
    P_\Psi(abc|xyz)=\tr(\proj{\Psi} M_x^a\otimes M_y^b\otimes M_z^c)
    \,,
\end{equation}
are genuine multipartite nonlocal, \ie violate a Svetlichny-like Bell inequality~\cite{svetlichny,svetlichnyN,collins} for the detection of genuine multipartite nonlocality in the no-signalling framework (see discussion at the end of Appendix A).

Given the $L$ copies of $\sigma$, the parties apply the following measurement strategies. Alice measures the flag $A_f$ of her first copy of $\sigma$ in the $\{\ket 0,\ket 1\}$ basis. Without loss of generality, assume her result is equal to 0. Then, she knows she shares a maximally-entangled state with Bob in the $AB$ system. She keeps measuring the remaining flags until she gets outcome 1. Then, she also shares a maximally-entangled state with Charlie in the $AC$ systems. That is, Alice has effectively prepared two singlets in a $\Lambda$ configuration. She now applies the measurement $M_x^a$, given in~\eqref{prob_Psi}, to the two particles associated to the two flags where she got first results 0 and 1. Bob and Charlie apply a similar strategy: Bob (Charlie) measures the
flags $B_f$ ($C_f$) in the computational basis until he finds result 0 (1). Then he applies the measurement $M_y^b$ ($M_z^c$) to the corresponding $B$ ($C$) particle. Although presented in a sequential way for clarity reasons, this process defines in fact one-shot local measurements on each
party. Clearly, the obtained correlations among the parties are the same as in
\eqref{prob_Psi} and, thus, are genuine tripartite nonlocal.

Of course, with probability $p_\mathrm{eq}=1/2^{L-1}$ all the flags give the same result. As it becomes clear in the next lines, these instances can be ignored, but we discuss them here for the sake of completeness. Assume that all the flags give 0 (1). Then, the parties can apply, for instance, the measurement strategy on $L$
maximally-entangled states between Alice and Bob (Charlie) which optimally approximates
$P_\Psi(abc|xyz)$. The obtained correlations
are denoted by $P_\Psi^B(abc|xyz)$ ($P_\Psi^C(abc|xyz)$), although their explicit form
is irrelevant for our considerations.

Putting all these
possibilities together, the resulting probability distribution among the three parties
is equal to
\begin{equation}\label{prapprox}
    \frac{p_\mathrm{eq}}{2} \tilde P_\Psi^B(abc|xyz) + \frac{p_\mathrm{eq}}{2} \tilde P_\Psi^C(abc|xyz) + (1-p_{eq})P_\Psi(abc|xyz) .
\end{equation}
This can be made arbitrarily close to the tripartite nonlocal distribution $P_\Psi(abc|xyz)$, as
$p_\mathrm{eq}$ tends to zero exponentially with the number of copies
$L$. Therefore, there always exists a finite value of $L$ (that might
not be large), such that genuine tripartite nonlocal correlations
can be obtained from $\sigma^{\otimes L}$, although $\sigma$ was not genuine tripartite nonlocal.

The example of activation of standard nonlocality can be obtained following a similar procedure.
As above, we combine classically correlated flags with the activation result in the network scenario to build an $N+1$-partite
quantum state $\tau$ which is local, but such that $\tau^{\otimes L}$ is nonlocal for
large $L$. The state $\tau$ reads
\begin{multline}\label{tau} \tau_{AB_1 B_2...B_N}=\\=\frac{1}{\sqrt
N}\sum_{i=1}^N \rho_{AB_i}\otimes
\ket{i_{A_f}i_{B_{1,f}}...i_{B_{N,f}}}\!\bra{i_{A_f}i_{B_{1,f}}...i_{B_{N,f}}},
\end{multline}
where
\begin{equation}
\rho_{AB_i}=\rho_{\mathrm{I}}^{AB_i}\bigotimes_{j=1,j\neq i}^N
\gamma^{B_j}.
\end{equation}
represents the product of a two-qubit isotropic state
$\rho_{\mathrm{I}}^{AB_i}$, shared by parties $A$ and $B_i$, with an
arbitrary state $\gamma_{B_j}$ for the remaining ones. The states $\ket{i_{A_f}}$ and $\ket{i_{B_{j,f}}}$ provide the correlated flags among the parties. 
Using a similar
measurement scheme as above, it is easy to prove that by increasing the
number of copies $L$, $\tau^{\otimes L}$ can be deterministically
transformed into a state arbitrarily close to the nonlocal
star-configuration state of Fig.\ref{star}.

\subsection{Any two-outcome measurements can be simulated by von Neumann measurements.}

Here we show that projective measurements are enough to simulate
any outcome distribution obtained by dichotomic (\ie two-outcome)
general measurements.

Consider a dichotomic measurement described by elements $M_0$ and $M_1$, which are positive operators such that $M_0+M_1=1$.
Then, their spectral decompositions can be expressed in the same basis: $M_0=\sum_i \lambda_i
\ket{\varphi_i}\bra{\varphi_i}$ and $M_1=\sum_i (1-\lambda_i)
\ket{\varphi_i}\bra{\varphi_i}$, with $0\leq\lambda_i\leq 1$. Consequently, the
results of this general measurement can be simulated by a protocol consisting of the following steps:
(i) the von Neumann measurement defined by projectors
$\{\ket{\varphi_i}\bra{\varphi_i}\}$ is applied and (ii) depending on the observed outcome, $i$, the observer outputs 0 with probability $\lambda_i$ and $1$ with probability $1-\lambda_i$.

This simple observation implies that a local model for projective
measurements  on a quantum state also applies to general
dichotomic measurements. This turns out to be particularly
relevant for our example of activation of nonlocality. Recall that
this example is based on the fact that the star-shape network made
of isotropic states, see Fig.~\ref{star}, is nonlocal for a noise
threshold $p>0.64$. It is however crucial that dichotomic
measurements are sufficient to reveal the nonlocality of the
network~\cite{sen}. While it is unknown whether there exist
isotropic states with $p>0.64$ that are local under general
measurements, as the best known model works for $p\leq
5/12$~\cite{barrett}, they are certainly local for two-outcome
measurements when $p<0.66$. This easily follows from the previous
result and the existence of a local model for projective
measurements when $p<0.66$~\cite{acin-gisin-toner} (see
Fig.\ref{bounds}). We then conclude that isotropic states have
their non-locality activated: there exist states local under
dichotomic measurements at the single-copy level, but which have
their nonlocality revealed by dichotomic measurements in the
network scenario.

\clearpage

\end{document}